\documentclass[pss]{wiley2sp} % provides pss two-column style
\usepackage{amsmath}
\begin{document}

% Title of the article
\title{Intrinsic spin Hall effect  in silicene: transition from spin Hall to normal insulator}

\titlerunning{Intrinsic spin Hall effect in silicene}

\author{%
  A. Dyrda\l\textsuperscript{\Ast,\textsf{\bfseries 1}},
  J. Barna\'s\textsuperscript{\textsf{\bfseries 1,2}},
  }

\authorrunning{A. Dyrda\l, J. Barna\'s}

\mail{e-mail
  \textsf{adyrdal@amu.edu.pl}, Phone:
  +48-61-8296 396, Fax: +48-61-8295 298}

\institute{%
  \textsuperscript{1}\,Department of Physics, Adam Mickiewicz University, ul. Umultowska 85, 61-614 Pozna\'n, Poland \\
  \textsuperscript{2}\,Institute of Molecular Physics, Polish Academy of Sciences,
  ul. M. Smoluchowskiego 17, 60-179 Pozna\'n, Poland}

\received{XXXX, revised XXXX, accepted XXXX} % do not change, will be filled in by the publisher
\published{XXXX} % do not change, will be filled in by the publisher


\abstract{%
\abstcol{%
Intrinsic contribution to the spin Hall effect in a
two-dimensional silicene is considered theoretically within the
linear response theory and Green function formalism. When an
external voltage normal to the silicene plane is applied, the spin
Hall conductivity is shown to reveal a  transition from the spin
Hall insulator phase at  low
  }{voltages to the conventional insulator phase at higher voltages. This
transition resembles recently reported phase transition in a
bilayer graphene. The spin-orbit interaction responsible for this
transition in silicene is much stronger than in graphene, which
should make the transition observable experimentally.   }}

\maketitle

\section{Introduction}
There is currently a great interest in two-dimensional IVA
materials like graphene and silicene. The peculiar electronic
transport properties of these materials, which follow from their
unusual band structure, make them promising for applications in
future nanoelectronics. As the properties of graphene are already
rather well known~\cite{castro,abergel}, the silicene -- a two-dimensional
silicon lattice -- has been synthesized only very
recently~\cite{Lalmi2010} and attracts great attention both
theoretical and experimental~\cite{kara2012}.

Similarly to graphene, the silicene is a material with linear
electronic energy spectrum near the Fermi level. The intrinsic
spin-orbit interaction, however, opens an energy gap, which in
silicene is relatively wide. The corresponding intrinsic gap in
graphene is rather negligible and therefore the effects due to its
presence are not observable experimentally. The large energy gap
as well as compatibility with silicon-based conventional
electronics make the silicene a very interesting material -- also
for spin electronics.

In this paper we study the spin Hall effect~\cite{dyakonov,hirsch}
in silicene, induced by intrinsic spin-orbit interaction. This
contribution to the spin Hall conductivity follows from a
nontrivial trajectory of charge carriers in the momentum space due
to the spin-orbit contribution of a perfect crystal lattice to the
corresponding band structure, and therefore it is also called
topological spin Hall effect (for review of SHE see e.g. Refs.
\cite{engel,dyakonov2008}).

The buckled structure of silicene (see Fig.1) makes the
corresponding spin-orbit interaction rather strong. Moreover, in
additon to the usual intrinsic spin-orbit interaction, there is
also an intrinsic Rashba like term, and both terms contribute to
the spin Hall conductivity. For simplicity, we neglect here the
role of structural defects, assuming perfect crystalline
two-dimensional lattice. As in the case of graphene, the spin Hall
conductivity is shown to be finite when the Fermi level is in the
gap. Since the Rashba type interaction is rather small, the
intrinsic spin Hall conductivity in silicene is almost quantized.
We also show that electric field (gate voltage) normal to the
silicene plane leads to a transition from the spin Hall insulator
phase to the normal insulator state, in agreement with predictions
by Ezawa~\cite{ezawa2012a,ezawa2012b}. We also note that the
transverse spin current in systems with spin-orbit coupling may be
generated by a temperature gradient instead of electric field, and
the effect is then known as thermal spin Hall or spin Nernst
effect.

\section{Model}

The crystallographic structure of silicene is shown schematically
in Fig.1 and is similar to graphene except that the planes
corresponding to the two atomic sublattices are now spatially
separated. The buckled atomic arrangement has two important
consequences. First, such a structure gives rise to a relatively
large intrinsic spin-orbit coupling responsible for a sizable
energy gap. This spin-orbit coupling also includes the Rashba like
term. Second, since the planes of the corresponding two
sublattices are separated, by applying perpendicular voltage one
can modify and control the electronic band structure. In the
following considerations we calculate the spin Hall conductivity
taking into account both terms in the intrinsic spin-orbit
coupling as well as the effects due to perpendicular external
voltage.

\begin{figure}[t]%
\includegraphics[width=0.9\linewidth,height=0.8\linewidth]{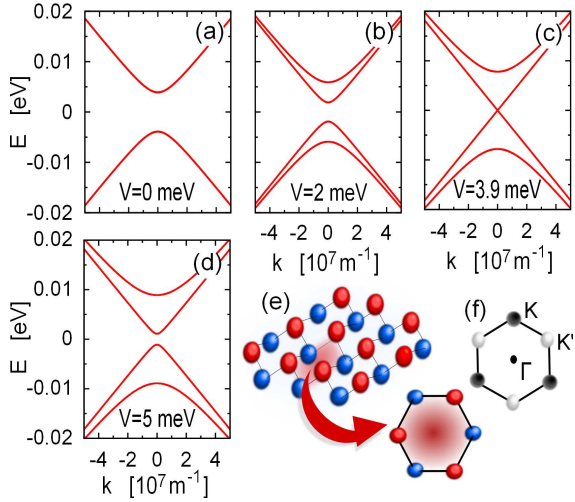}
\caption{Atomic and electronic band structure of silicene.
Electronic states near the energy gap for indicated gate voltages
(a-d), two dimensional crystallographic structure of silicene (e)
and the corresponding Brillouin zone (f).} \label{onecolumnfigure}
\end{figure}

To describe electronic states in silicene we use an effective
Hamiltonian that was derived within the tight-binding method and
first-principle calculations~\cite{Liu2011}. In the sublattice
space, the matrix form of this Hamiltonian for the K point of the
Brillouin zone takes the form
\begin{equation}
H_{0} = \left(
          \begin{array}{cc}
           h  & v (k_{x} + i k_{y}) \\
            v (k_{x} - i k_{y}) & - h \\
          \end{array}
        \right),
\end{equation}
where $v = \hbar v_{F}$ with $v_{F}$ being the Fermi velocity, and
\begin{equation}
h = - \lambda_{so} \sigma_{z} - a \lambda_{R} (k_{y} \sigma_{x} -
k_{x} \sigma_{y}) + V \sigma_{0}.
\end{equation}
The parameters $\lambda_{so}$ and $\lambda_{R}$ describe the two
terms of the intrinsic spin-orbit interaction, $V$ is the
perpendicular voltage (in energy units), and $\sigma_\alpha$ are
the unit ($\alpha =0$) and Pauli ($\alpha =x,y,z$) matrices in the
spin space. The corresponding energy spectrum near the K point is
shown in Fig.1(a-d) for indicated gate voltages. It is evident
that gate voltage first closes the gap and then opens it again
when $V$ increases further.

Density of spin current in the silicene plane can be  written as
$J_{i}^{s_{n}} = \sum_{j} \sigma^{s_{n}}_{ij} E_{j}$ (for
$i,j=x,y$), where $\sigma^{s_{n}}_{ij}$ is the spin conductivity
with $s_n=(\hbar /2)\sigma_n$ being the $n$-th component
($n=x,y,z$) of electron spin, and $E_{j}$ is the $j$th component
of the in-plane electric field. The quantum-mechanical operator of
spin current density may be defined as
$\textbf{J}^{s_{n}}=(1/2)\left[\textbf{v},s_{n}\right]_{+}$, where
$v_i=(1/\hbar )(\partial H/\partial k_i)$ is the velocity operator
($i=x,y$). In the linear response theory, the zero-frequency spin
Hall conductivity is given by the formula~\cite{dyrdal},
\begin{eqnarray}
\label{sigom} \sigma^{s_{z}}_{xy}(\omega)=\lim_{\omega\to
0}\frac{e \,\hbar }{2\omega}{\rm
Tr}\int\frac{d\varepsilon}{2\pi}\frac{d^{2}\mathbf{k}}{(2\pi)^{2}}
\left[v_{x},s_{z}\right]_{+}
\nonumber \\
G_{\mathbf{k}}(\varepsilon +
\omega)v_{y}G_{\mathbf{k}}(\varepsilon),
\end{eqnarray}
where $G_{\mathbf{k}}(\varepsilon )$ is the Green function
corresponding to the Hamiltonian (1) of the system.

\section{Spin Hall conductivity}

We consider now spin Hall conductivity, separately for the case
without vertical bias and for the case with a vertical voltage
applied. Let us begin with the former case.

\paragraph{The case of $V = 0$}

Calculating the appropriate Green functions and following the
general procedure \cite{dyrdal}, one finds the spin Hall
conductivity from Eq.(3). When the Fermi level $\mu$ (measured fom
the middle of the gap) is located either in the valence or
conduction bands, the spin Hall conductivity is given by
\begin{equation}
\sigma^{s_{z}}_{xy} = - \frac{v^{2} \lambda_{so}}{{\vert \mu
\vert} (v^{2} + a^{2} \lambda^{2}_{R})} \frac{e}{4 \pi}.
\end{equation}
When, in turn,  the Fermi level is inside the energy gap, one
finds
\begin{equation}
\sigma^{s_{z}}_{xy} = - \frac{v^{2}}{v^{2} + a^{2}
\lambda^{2}_{R}}\frac{e}{4 \pi}.
\end{equation}
One can note, that due to the intrinsic Rashba type spin-orbit
coupling, the spin Hall conductivity for the Fermi level in the
energy gap is generally not quantized. However, the parameter
$\lambda_{R} = 0.7$ meV in silicene is relatively small, so the
main contribution to the spin-orbit coupling is due to the term
proportional to $\lambda_{so}$. Indeed, assuming the relevant
parameters one finds $\sigma^{s_{z}}_{xy}(\lambda_{R} = 0) -
\sigma^{s_{z}}_{xy}(\lambda_{R} \neq 0) \approx 10^{-7}$. Thus,
the influence of intrinsic Rashba spin-orbit interaction on the
spin Hall conductivity is rather minor and will be neglected in
the following.  Accordingly, assuming  $\lambda_{R} = 0$ one may
rewrite Eqs (4) and (5) as
\begin{equation} \sigma^{s_{z}}_{xy} = -
\frac{\lambda_{so}}{\vert \mu\vert  } \frac{e}{4 \pi},
\end{equation}
\begin{equation}
\sigma^{s_{z}}_{xy} = - \frac{e}{4 \pi},
\end{equation}
for the Fermi level in the conduction/valenece bands and in the
energy gap, respectively. The latter formula shows that spin Hall
effect in silicene is quantized when the Fermi level is in the
gap.

\subsection{SHE with biased voltage}

When the external vertical voltage is applied, one finds
\begin{equation}
\sigma^{s_{z}}_{xy} = - \frac{\lambda_{so}}{\vert\mu\vert}
\frac{e}{4 \pi},
\end{equation}
if the Fermi level crosses both valence or both conduction bands,
$|\mu|
> |V| + \lambda_{so}$.
In turn, the spin Hall conductivity for $|V| + \lambda_{so} >
|\mu|
> ||V| - \lambda_{so}|$ is given by
\begin{equation}
\sigma^{s_{z}}_{xy} = - \frac{1}{2} \left( 1 - \frac{|V| -
\lambda_{so}}{\vert\mu\vert}\right)\frac{e}{4 \pi},
\end{equation}
while for the Fermi level inside the energy gap, $|\mu| < ||V| -
\lambda_{so}|$, one finds
\begin{equation}
\sigma^{s_{z}}_{xy} = - \frac{e}{4 \pi}.
\end{equation}

\begin{figure}[t]%
\includegraphics[width=0.9\linewidth,height=0.65\linewidth]{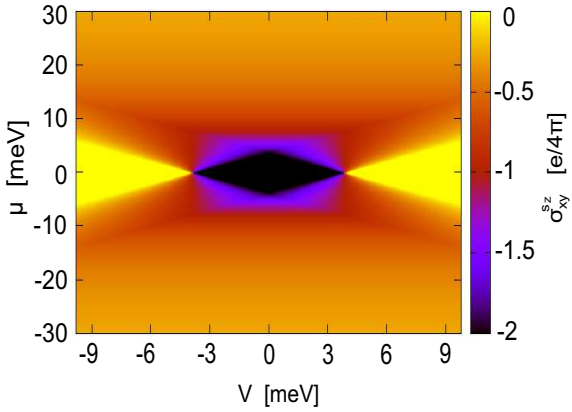}
\caption{Spin Hall conductivity as a function of the vertical
voltage $V$ and the Fermi level position. Contributions from both
$K$ points are included. The other parameters are: $v = \hbar
v_{F}$ with $v_{F} = 5.52 \cdot 10^{5}$ m/s, and $\lambda_{so} =
3.9$ meV.} \label{fig1}
\end{figure}

\begin{figure}[h]%
\includegraphics[width=0.9\linewidth,height=0.65\linewidth]{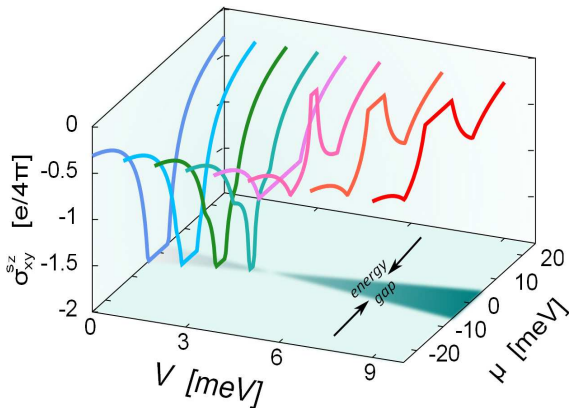}
\caption{Some cross-sections of Fig.\ref{fig1} for selected values
of $V$.} \label{onecolumnfigure}
\end{figure}

Behavior of the spin Hall conductivity with increasing voltage $V$
and position of the Fermi level $\mu$ is shown in Fig.2, where
contributions from both $K$ points of the Brillouin zone are
included. This figure clearly shows that the energy gap, where the
spin Hall conductivity is finite and quantized, shrinks with
increasing voltage $|V|$. At some critical value of $|V|$ the gap
becomes closed, and then opens again with $|V|$ increasing
further. However, the spin Hall conductivity vanishes then for
$\mu$ in the gap. This behavior is also evident in Fig.3, where
some cross-sections of Fig.2 for selected values of the voltage
$V$ are plotted. Both figures clearly indicate a transition from
spin Hall insulator phase at voltages $|V|$ smaller than the
critical voltage to conventional insulating behavior at voltages
$|V|$ larger than the critical one.

\section{Summary}

To summarize, we have derived some analytical formula for spin
Hall conductivity  in silicene due to intrinsic spin-orbit
interaction. We have shown that the intrinsic Rashba type term in
spin-orbit interaction has a negligible influence on the spin Hall
conductivity, so the latter is quantized when the Fermi level is
inside the gap. Furthermore, we have shown that there is a
transition from the spin Hall insulator phase to the conventional
insulator one with increasing external gate voltage applied
perpendicularly to the layer plane.

\vspace{-0.2cm}
\begin{acknowledgement}
This work has been supported by the European Union under European
Social Fund - Operational Programme  'Human Capital' -
POKL.04.01.01-00-133/09-00.
\end{acknowledgement}


\begin{thebibliography}{[1]}

\bibitem{castro}
A.\,H.~Castro Neto, F.~Guinea, N.\,M.\,R.~Pers, K.\,S.~Novoselov, and A.\,K.~Geim,
Rev. Mod. Phys. {\bf 81}, 109 (2009).

\bibitem{abergel}
D.\,S.\,L.\, Abergel, V. Apalkov, J. Berashevich, K. Ziegler, and T. Chakraborty,
Advances in Physics, {\bf 59}, No. 4, 261 (2010).

\bibitem{Lalmi2010}
B. Lalmi, H. Oughaddou, H. Enriquez, A. Kara, S. Vizzini, B. Ealet, and B. Aufray,
Appl. Phys. Lett. \textbf{97}, 223109 (2010).

\bibitem{kara2012}
A. Kara, H. Enriquez, A.\,P. Seitsonen,
L.\,C.\,L.\,Y. Voon, S. Vizzini, B. Aufray, H. Oughaddou,
Surface Science Reports \textbf{67}, 1-18 (2012).

\bibitem{dyakonov}
M.\,I.~Dyakonov, V.\,I.~Perel, Pis'ma Z. Eksp. Teor. Fiz. {\bf 13},
657 (1971); JETP Lett. {\bf 13}, 467 (1971).

\bibitem{hirsch}
J.~E.~Hirsch, Phys. Rev. Lett. {\bf 83}, 1834 (1999).

\bibitem{engel}
H.~A.~Engel, E.~I.~Rashba, B.~I.~Halperin, in {\em Handbook of
Magnetism and Advanced Magnetic Materials}, edited by
H.~Kronmuller, S.~Parkin, vol.~5: {\em Spintronics and
Magnetoelectronics}, John Willey, New York, 2007.

\bibitem{dyakonov2008}
M.~I.\,Dyakonov and A.~V.\,Khaetskii,
in {\em Spin Physics in Semiconductors} edited by
M.~I.\,Dyakonov, Chap. 8, Springer-Verlag, Berlin,
Heidelberg,  2008.

\bibitem{ezawa2012a}
M.\,Ezawa, arXiv:1201.3687v1 (2012).

\bibitem{ezawa2012b}
M.\,Ezawa, arXiv:1202.1357 (2012).


\bibitem{Liu2011}%
 C.-C. Liu, H. Jiang, and Y. Yao,
 Phys. Rev. B \textbf{84}, 195430 (2011).


\bibitem{dyrdal}
A. Dyrda\l, V.K. Dugaev, J. Barna\'s,
Europhysics Letters \textbf{85}, 67004 (2009)


\end{thebibliography}
\end{document}